\documentclass[twocolumn,english,showkeys,showpacs,superscriptaddress]{revtex4-1}
\usepackage[LGR,T1]{fontenc}
\usepackage[latin9]{inputenc}
\usepackage{amsmath}
\usepackage{amssymb}
\usepackage{graphicx}

\makeatletter

\ProvideTextCommand{\~}{LGR}[1]{\char126#1}

\usepackage[normalem]{ulem}

\usepackage{color}

\usepackage{ulem}

\newcommand{\stkout}[1]{\ifmmode\text{\sout{\ensuremath{#1}}}\else\sout{#1}\fi}

\makeatother

\usepackage{babel}
\begin{document}
\title{ Frequency Measurement with Superradiant Pulses of Incoherently Pumped Calcium Atoms: Role of Quantum Measurement Backaction}

\author{Huihui Yu}
\address{Henan Key Laboratory of Diamond Optoelectronic Materials and Devices, Key Laboratory of Material Physics Ministry of Education, School of Physics and Microelectronics, Zhengzhou University, Daxue Road 75, Zhengzhou 450052, China}

\author{Yuan Zhang}
\email{yzhuaudipc@zzu.edu.cn}
\address{Henan Key Laboratory of Diamond Optoelectronic Materials and Devices, Key Laboratory of Material Physics Ministry of Education, School of Physics and Microelectronics, Zhengzhou University, Daxue Road 75, Zhengzhou 450052, China}

\author{Qilong Wu}
\address{Henan Key Laboratory of Diamond Optoelectronic Materials and Devices, Key Laboratory of Material Physics Ministry of Education, School of Physics and Microelectronics, Zhengzhou University, Daxue Road 75, Zhengzhou 450052, China}




\author{Chongxin Shan}
\email{cxshan@zzu.edu.cn}
\address{Henan Key Laboratory of Diamond Optoelectronic Materials and Devices, Key Laboratory of Material Physics Ministry of Education, School of Physics and Microelectronics, Zhengzhou University, Daxue Road 75, Zhengzhou 450052, China}

\author{Klaus M{\o}lmer}
\email{klaus.molmer@nbi.ku.dk}
\address{Niels Bohr Institute, University of Copenhagen, Blegdamsvej 17, 2100 Copenhagen, Denmark}

\begin{abstract}
A recent experiment demonstrated heterodyne detection-based frequency measurements with superradiant pulses from coherently pumped strontium atoms in an optical lattice clock system, while another experiment has analyzed the statistics of  superradiant pulses from incoherently pumped calcium atoms in a similar system. In this article, we propose to perform heterodyne detection of the superradiant pulses from the calcium atoms, and analyze theoretically the corresponding atomic ensemble dynamics in terms of the rotation of a collective spin vector and the incoherent quantum jumps among superradiant Dicke states. We examine the effect of quantum measurement backaction on the emitted field and the collective spin vector dynamics, and we demonstrate that it plays an essential role in the modelling of  the frequency measurements. We develop a stochastic mean field theory, which is also applicable to model frequency measurements with steady-state superradiance signals,  and  to explore quantum measurement effects in the dynamics of atomic ensembles.
\end{abstract}

\maketitle



\section{Introduction}

Atomic clocks are used for international time standards, national time services, satellite navigation systems~\citep{BJaduszliwer}, and studies of general relativity ~\citep{CHafeleJ}. By exploring optical ($\sim  10^5$ GHz) rather than microwave ($\sim {\rm GHz}$) transitions ~\citep{ABauch}, the performance of the clocks improve by orders of magnitude, and optical clocks are thus subject of  considerable attention ~\citep{LudlowAD}. In particular, optical lattice clocks are now extensively pursued because they mitigate the Doppler effect  and provide better signal-to-noise ratio with more atoms ~\citep{HKatori}.

The frequency of atomic clocks is usually measured passively by locking a laser to the atomic transition and maximizing the fluorescence signal. This passive procedure has good long-term stability and absolute accuracy. Alternatively, the clock frequency can be measured actively by comparing the radiation from the atoms with a reference laser. This active approach offers a wide detection bandwidth and dynamical range and has been applied with hydrogen masers in the microwave domain~\citep{MAWeiss}. Recently, it was applied on optical superradiance pulses from coherently driven strontium atoms ~\citep{MANorcia2018}, and a stochastic mean-field theory for such frequency measurements has been developed to reveal the physical mechanisms involved~\citep{YZhang2021-1}.

\begin{figure}[!htp]
    \centering
    \includegraphics[width=0.42\textwidth]{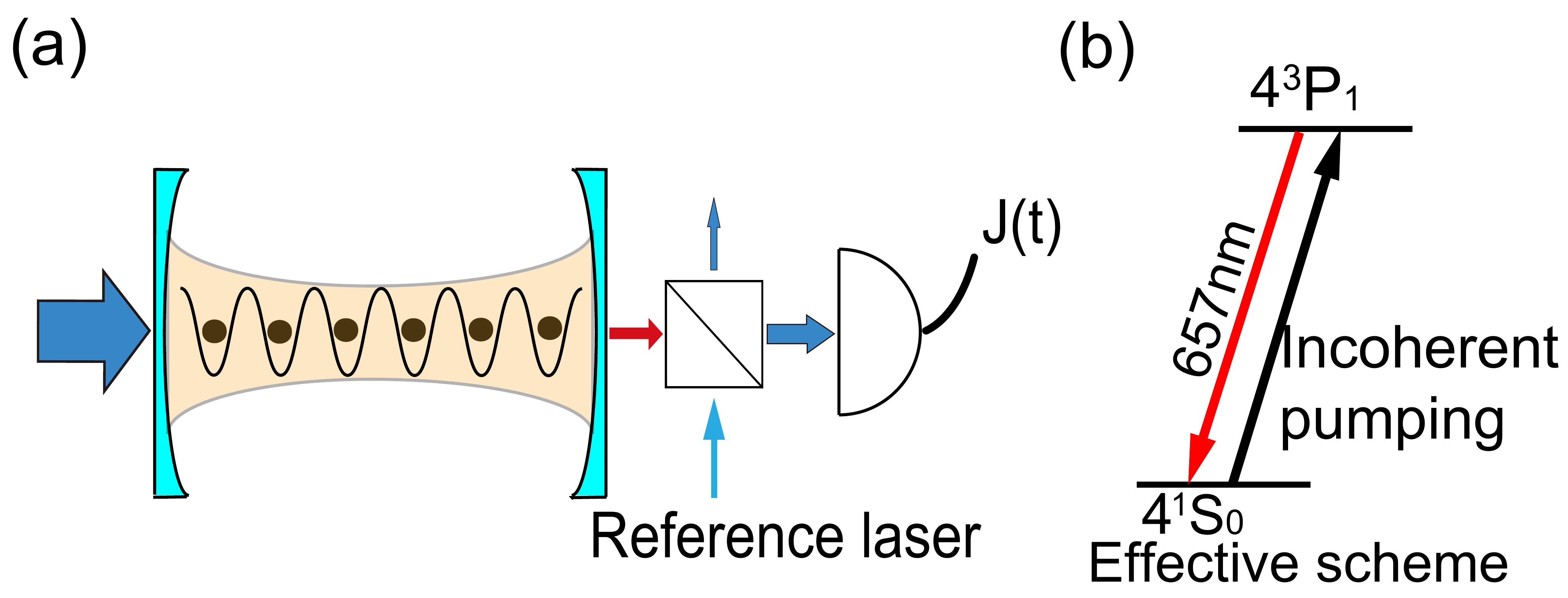}
    \caption{System and energy diagram. Panel (a) shows tens of thousands of calcium atoms trapped in an optical lattice inside an optical cavity, and the proposed heterodyne detection of the emitted superradiant pulses from the incoherently pumped atoms. Panel (b) shows the simplified energy scheme with the atoms incoherently pumped to the $4^3{\rm P}_1$ state (black arrow), and the Purcell-enhanced and collective decay of the excited atoms to the electronic ground $4^1{\rm S}_0$ via the optical cavity mode (red arrow). For more details, see the text. }
    \label{fig:sys}
\end{figure}

In Ref.~\citep{TLaske}, Laske et al.  reported the observation of superradiant pulses emitted on the narrow-band inter-combination line ${\rm 4^{3}P_1} - {\rm 4^{1}S_0}$ of calcium atoms, trapped inside a resonant optical cavity [Fig. \ref{fig:sys}(a)]. There, the calcium atoms were initially prepared on the excited state via a complex process involving a MOT and an optical lattice trap~\citep{DPHansen}   [see Fig. \ref{fig:scheme}], which can be modeled effectively by an incoherent pumping rate~\citep{DMeiser2009,KDebnath2018,CHotter} (right part of Fig. \ref{fig:sys}a). This incoherent pumping mechanism differs from the excitation scheme used in strontium experiments~\citep{MANorcia2018}, and this affects the theoretical description of the superadiance process. 

In this article, we study in detail how the superradiance pulses of the incoherently pumped calcium atoms depend on the number of atoms, the pumping duration and excitation rate. Our study reveals that the superradiance pulses are here caused by incoherent but collectively enhanced decay among the Dicke states, which is in strong contrast to the superradiance from the coherently driven strontium atoms caused by the rotation of a collective spin vector. Furthermore, we propose to carry out the heterodyne detection-based frequency measurement with the superradiance from the calcium atoms, and we show that the quantum measurement backaction plays an essential role, and leads to shot-to-shot statistics of the superradiant pulses. Precision frequency measurements thus constitute an excellent example for exploration and application of quantum measurement effects on atomic ensemble system, which will be also at play in the account of frequency measurements with steady-state superradiance~\citep{DMeiser2009,JDBohnet2012}. 


\section{Superradiant Pulses from Incoherently Pumped Atoms}


To investigate superradiant pulses from $N$ incoherently pumped calcium atoms, we consider the quantum master equation $\partial_t \hat{\rho} =-\frac{i}{\hbar}\left[\hat{H}_{c}+\hat{H}_{a}+\hat{H}_{a-c},\hat{\rho}\right] -\kappa\mathcal{D}\left[\hat{a}\right]\hat{\rho}-\gamma\sum_{k=1}^{N}\mathcal{D}\left[\hat{\sigma}_{k}^{12}\right]\hat{\rho} -\eta\sum_{k=1}^{N}\mathcal{D}\left[\hat{\sigma}_{k}^{21}\right]\hat{\rho}-(\chi/2)\sum_{k=1}^{N}\mathcal{D}\left[\hat{\sigma}_{k}^{22}-\hat{\sigma}_{k}^{11}\right]\hat{\rho}$ for the reduced density operator $\hat{\rho}$. The Hamiltonian $\hat{H}_{c}=\hbar\omega_{c}\hat{a}^{\dagger}\hat{a}$
describes the optical cavity field with frequency $\omega_{c}$ and photon creation $\hat{a}^{\dagger}$ and annihilation operator $\hat{a}$. The Hamiltonian $\hat{H}_{a}=\hbar\omega_{a}\sum_{k=1}^{N}\hat{\sigma}_{k}^{22}$ describes the atomic ensemble with atoms labeled by $k$ and the projection operators $\hat{\sigma}_{k}^{11(22)}$ on the ground state ${\rm 4^{1}S_0}$ (excited state ${\rm 4^{3}P_1}$). The interaction Hamiltonian $\hat{H}_{a-c}=\hbar g\sum_{k=1}^{N}\left(\hat{a}^{\dagger}\hat{\sigma}_{k}^{12}+\hat{\sigma}_{k}^{21}\hat{a}\right)$ describes the atom-cavity coupling with a strength $g$, where $\hat{\sigma}_{k}^{12},\hat{\sigma}_{k}^{21}$ are the lowering and raising operator of the atoms. The Lindblad terms describe dissipation processes by the superoperators $\mathcal{D}\left[\hat{o}\right]\hat{\rho}=\frac{1}{2}\left(\hat{o}^{\dagger}\hat{o}\hat{\rho}+\hat{\rho}\hat{o}^{\dagger}\hat{o}\right)-\hat{o}\hat{\rho}\hat{o}^{\dagger}$
(for any operator $\hat{o}$), where $\kappa$ is the loss rate of the intra-cavity photons and $\gamma,\eta,\chi$ are the spontaneous emission rate, the incoherent pumping rate and the dephasing rate of the atoms, respectively.

We employ the QuantumCumulants.jl package~\citep{DPlankensteiner} to solve the quantum master equation within the second-order mean-field approach. Further,  we assume identical parameters for all atoms and explore the permutation symmetry of the state to reduce dramatically the number of independent equations~\citep{DMeiser2009}, see Fig. \ref{fig:code} and Fig. \ref{fig:mf_det_sto} for the relevant codes. As a result, we obtain equations for mean-field quantities, such as the intra-cavity field amplitude $\left \langle \hat{a}^\dagger \right \rangle$ and atomic coherences $\left \langle \hat{\sigma}_{1}^{12} \right \rangle$, the intra-cavity mean photon number $\left \langle \hat{a}^\dagger \hat{a} \right \rangle$ and mean population of the excited atomic state $\left \langle \hat{\sigma}_{1}^{22} \right \rangle$. Importantly, we also obtain equations for the atom-field and atom-atom correlations $\left \langle \hat{a}^\dagger\hat{\sigma}_{1}^{12} \right \rangle$, $\left\langle \hat{\sigma}_{1}^{21}\hat{\sigma}_{2}^{12}\right\rangle, \left\langle \hat{\sigma}_{1}^{22}\hat{\sigma}_{2}^{22}\right\rangle$. 

To visualize the dynamics of the atomic ensemble, we follow the description in our previous works~\citep{KDebnath2018,YZhang2021-1,YZhang2021} to employ the Dicke states $\left | J, M \right \rangle$, where the integer or half-integers $J,M$ in the range $J\le N/2,-J\le M\le J$ describe the degree of symmetry and the excitation of the ensemble of two-level systems~\citep{RHDicke1954}, respectively. The Dicke states for given $J$ can be visualized as vertical ladders with equal spacing between neighbouring values of $M$, while the ladders for different $J$ are shifted horizontally to form a triangular shaped discrete set of states~\citep{LMandel}. The optical cavity-mediated radiation rate of the atomic ensemble  depends on the number of photons and the quantum numbers  $J,M$ in a non-trivial way while, generally, the Dicke states with larger and smaller $J$ have faster and slower decay rate, and are thus identified as superradiative and sub-radiative states. The influence of the various processes in the master equation can be conveniently visualized in the space of Dicke states~\citep{YZhang4,NShammah}, and here we shall employ this space to represent the results obtained by the mean-field method. To this end, we use the following expressions to calculate the averages of the Dicke state quantum numbers $\overline{M}=N\left(\left\langle \hat{\sigma}_{1}^{22}\right\rangle -\frac{1}{2}\right)$ and   $\overline{J}=\sqrt{\frac{3}{4}N+N\left(N-1\right)\left(\left\langle \hat{\sigma}_{1}^{21}\hat{\sigma}_{2}^{12}\right\rangle +\left\langle \hat{\sigma}_{1}^{22}\hat{\sigma}_{2}^{22}\right\rangle -\left\langle \hat{\sigma}_{1}^{22}\right\rangle +\frac{1}{4}\right)}$~\citep{KDebnath2018,YZhang2021}. 

The length of the collective Dicke spin depends explicitly on the atom-atom coherences and correlations, but does not assume a non-vanishing  (single atom) mean optical coherence. Such a coherence, however, plays a crucial role both in the single atom and in the collective spin vector of the atomic ensemble $\mathbf{A}=\sum_{i=x,y,z} A_i\mathbf{e}_i$, where $A_x = (N/2)(\langle \hat{\sigma}^{12}_1 \rangle + \langle \hat{\sigma}^{21}_1\rangle), 
A_y = (iN/2)( \langle\hat{\sigma}^{12}_1\rangle-\langle\hat{\sigma}^{21}_1\rangle),
A_z = (N/2)(\langle\hat{\sigma}^{22}_1\rangle -\langle\hat{\sigma}^{11}_1\rangle)= \overline{M}$ and $\mathbf{e}_i$ are the unit vectors  of the Cartesian coordinate system.  For a pure product state of the atoms, the collective spin vector traces a sphere of radius $N/2$, where the south (north) pole represents the atomic ensemble in the ground (excited) state. For mixed and entangled atomic states, the collective spin vector dynamics occurs inside this sphere. 

\begin{figure}[!htp]
\centering
\includegraphics[width=0.48\textwidth]{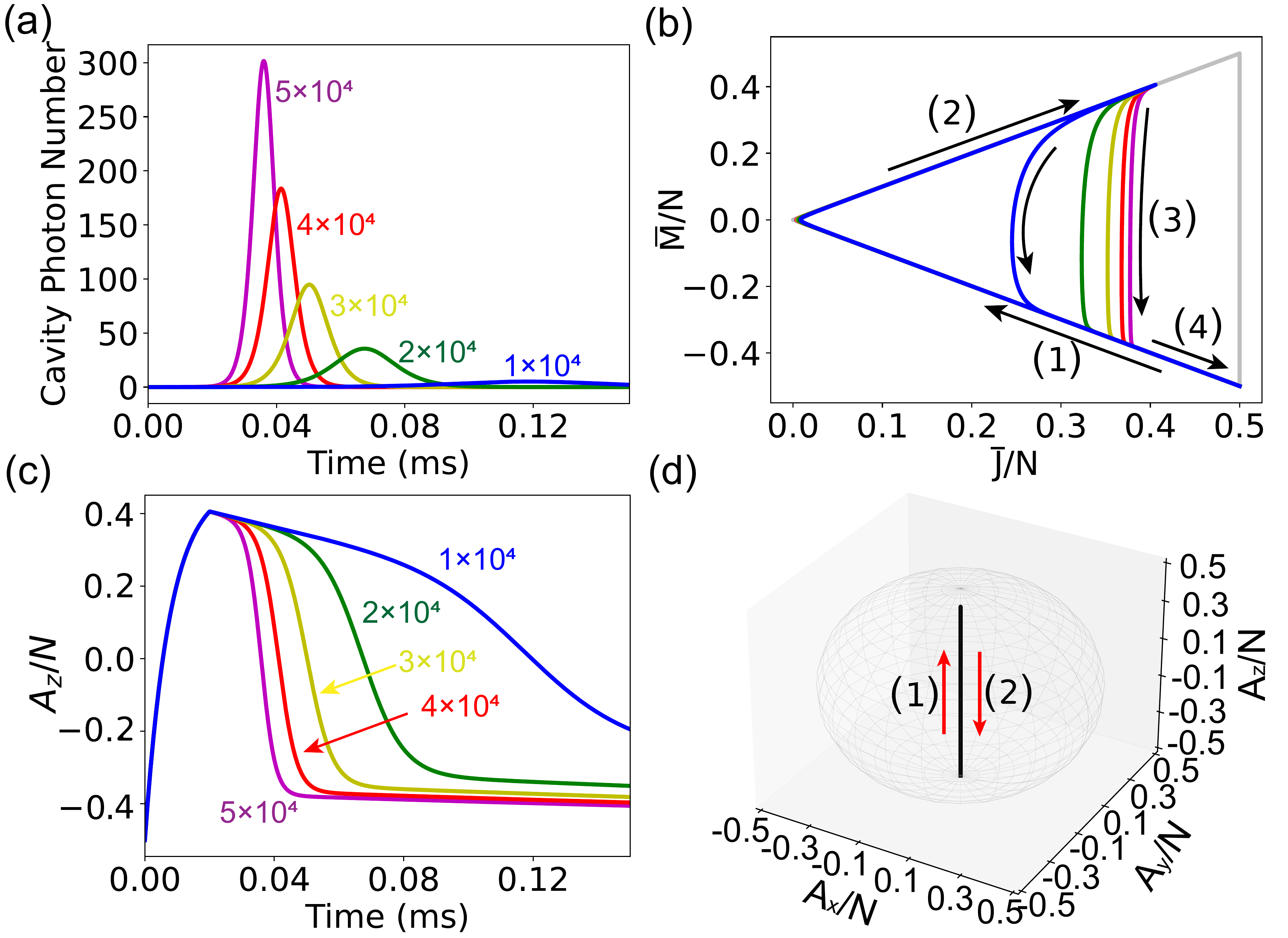}
\caption{\label{fig:pulses}Superradiant pulses from incoherently pumped calcium atoms. Panel (a,b) show as function of time the intra-cavity photon number (a) and the averaged Dicke state quantum numbers  $\overline{J},\overline{M}$ (b), which start from the ground state, i.e., the lower right corner, and follow the four enumerated arrows, for systems with  $N=10^4,2\times 10^4, 3\times 10^4, 4\times 10^4, 5\times 10^4$ atoms and the incoherent pumping pulse of rate $20$ kHz and duration $20$ $\mu$s. Panel (c,d) show the dynamics of the collective spin vector component $A_z$ (equivalent to $\overline{M}$) for different numbers of atoms (c), and
the dynamics of the collective spin vector for $N=10^4$ atoms (d), indicating the collective atomic inversion and (vanishing) coherences. In the panels (b-d), all the quantities are normalized by the number of atoms $N$. 
Other parameters are given in Tab.~\ref{tab:params}.} 
\end{figure}

We have solved the master equation within the cumulant mean-field theory with the parameters of the experiment~\citep{TLaske}, and we show in Fig. \ref{fig:pulses} the results for systems with different number of atoms, including the dynamics of the intra-cavity photon number (a) and the atomic ensemble Dicke state quantum numbers (b) and the collective spin vector (c,d). Fig. \ref{fig:pulses}(a) shows that the intra-cavity photon number is initially very small, and then increases and decreases abruptly, forming a superradiant pulse. In calculations with fewer atoms, the center of the pulses shifts to longer times, the duration of the pulses increases, and the maximum number of photons decreases. These results agree well with the observations in the experiment~\citep{TLaske}. In particular, we note that in our simulations the cavity field amplitude initially vanishes and it does not develop finite values at any later time [see  Fig. \ref{fig:Cavity Field}(a)].  

Fig. \ref{fig:pulses}(b) shows that the ground state atomic ensemble occupies initially the Dicke states at the lower-right corner. During the incoherent pumping of the atoms, the atomic ensemble progresses with increasing $M$-values while exploring first lower and then higher $J$-values along the lower and higher boundary of the Dicke state space triangle. Finally, the ensemble decays collectively, preserving high permutation symmetry as it progresses vertically downwards the ladders to the states on the lower Dicke state boundary. Although the normalized value $\overline{J}/N$ after the incoherent pumping is approximately independent of the number of atoms, the collective, superradiant emission rate depends on $\overline{J}$ and thus increases with increasing $N$. Thus, the shorter delay of the superradiant pulses for larger $N$ is caused by the enhanced transition rates among the states on the upper boundary of Dicke state space after the incoherent pumping. The reduced pulse duration for large $N$ is due to the enhanced transition rates among the states with larger $\overline{J}$, and the increased peak intensity for larger $N$ is caused by the transition rates between the Dicke states with $\overline{M}\sim 0$, which increase quadratically with $\overline{J}$. For systems with fewer atoms, the atomic spontaneous decay rate becomes comparable with the collective decay rate and causes a relatively larger reduction of $\overline{J}$ values during the superradiant decay. Furthermore, in Fig. \ref{fig:pump time}, we investigate also the dependence of the superradiant pulses on the incoherent pumping rate and pumping time. We find that by increasing these parameters, we obtain stronger and shorter superradiant pulses  because we prepare the atoms in Dicke states with larger $\overline{J}$ and $\overline{M}$.

Fig. \ref{fig:pulses}(c) shows the collective spin vector component $A_z$ (normalized by the number of atoms $N$), which is equal to $\overline{M}$, as a function of time. We see that $A_z/N$ increases smoothly up to $0.4$ during the incoherent pumping, and then decays dramatically after a slow decline, and approaches a finite value of $\sim -0.4$, which is consistent with the lowest possible value of $\overline{M}$ for the given value of $\overline{J}$. For long time after the superradiant pulses, $A_z/N$ decreases slowly to $-0.5$, due to the much slower spontaneous emission of individual atoms, see Fig. \ref{fig:pump time}(f).  We observe that as the number of atoms increases, the decay occurs earlier and becomes faster. 
Due to the incoherent excitation, the atomic coherence, and hence, the other collective spin vector components $A_x,A_y$ vanish throughout the process. During the superradiant emission, the collective spin vector thus evolves along the z-axis from the point near the north pole through the origin to some points near the south pole as shown in Fig. \ref{fig:pulses}(d). This is in contrast to the dynamics in the superradiant pulses from coherently driven atoms~\citep{MANorcia2018}, where the collective spin vector rotates on the surface of the sphere, see also Fig. \ref{fig:Coherent pump}. 


While the vanishing of  $A_x,A_y$ is consistent with a vanishing mean value of the field amplitude $\left\langle \hat{a}^\dagger \right\rangle$, it does not imply the absence of superradiance, which is ultimately governed by the enhanced collective transition rates among the Dicke states. Although the superradiant power spectrum as often calculated with quantum regression theory does not require  a non-vanishing  cavity  field amplitude, the description of active frequency measurements by the theory of continuous quantum measurements, indeed, produces states with a non-vanishing cavity field amplitude, see more discussions below and in the Appendix~\ref{sec:MeaCoh}.

\section{Active frequency Measurement with Superradiant Pulses}

In the following, we describe heterodyne detection of the field emitted by the optical cavity by employing a stochastic master equation that supplements the deterministic equation $\partial_t \hat{\rho}$ with the stochastic measurement backaction~\citep{HWWiseman} $
\left(\partial_t \hat{\rho}\right)_{mea}=\frac{dW}{dt}\sqrt{\xi\kappa/2}\Bigl[e^{i\omega_{l}t}\left(\hat{a}-\left\langle \hat{a}\right\rangle \right)\hat{\rho}+\mathrm{h.c.}\Bigr]$, where $dW/dt$ describes  the photon shot-noise in the detector and the random
number $dW$ follows a normal distribution with a variance $dW^{2}=dt$
and a mean $E\left[dW\right]=0$. The parameter $\xi$ denotes the efficiency of the heterodyne detection, and $\omega_{l}$ is the local oscillator frequency. After subtraction of a constant contribution from the reference laser power, the current in the heterodyne detector is given by an expression on the form $J\left(t\right)=\sqrt{\xi\kappa/2}\mathrm{Re}\left[e^{-i\omega_{l}t}\left\langle \hat{a}^{\dagger}\right\rangle \left(t\right)\right]+dW/dt
$, with a term that is proportional to a quadrature of the intra-cavity field amplitude $\left\langle \hat{a}^{\dagger}\right\rangle$ and a term representing the photon shot-noise. 

We modify the QuantumCumulant.jl package to account for the contribution of $\left(\partial_t \hat{\rho}\right)_{mea}$ to the mean-field equation, see the Appendix B. The evolution of the mean field amplitude $\langle\hat{a}^{\dagger}\rangle$ is thus given by:
\begin{align}
 & \partial_{t}\langle\hat{a}^{\dagger}\rangle=\left(i\omega_{c}-\kappa/2\right)\langle\hat{a}^{\dagger}\rangle+iNg\langle\hat{\sigma}_{1}^{21}\rangle\nonumber \\
 & +\frac{dW}{dt}\sqrt{\xi\kappa/2}e^{i\omega_{l}t}\left(\langle\hat{a}^{\dagger}\hat{a}\rangle-\langle\hat{a}^{\dagger}\rangle\langle\hat{a}\rangle\right)\nonumber \\
 & +\frac{dW}{dt}\sqrt{\xi\kappa/2}e^{-i\omega_{l}t}\left(\langle\hat{a}^{\dagger}\hat{a}^{\dagger}\rangle-\langle\hat{a}^{\dagger}\rangle^{2}\right).\label{eq:field}
\end{align}
This equation indicates that the cavity field amplitude can develop a non-vanishing value from 
quantum measurement backaction. 
Although $\langle\hat{a}^{\dagger}\rangle$ and $A_x,A_y$ are strictly zero in the unconditioned average dynamics as analyzed in Fig.~\ref{fig:pulses}, they develop non-vanishing values due to the quantum measurement backaction of heterodyne detection. This is in contrast to the origin of these values in the coherent driving of strontium atoms~\citep{YZhang2021-1}. 

\begin{figure}[!htp]
    \centering
    \includegraphics[width=0.48\textwidth]{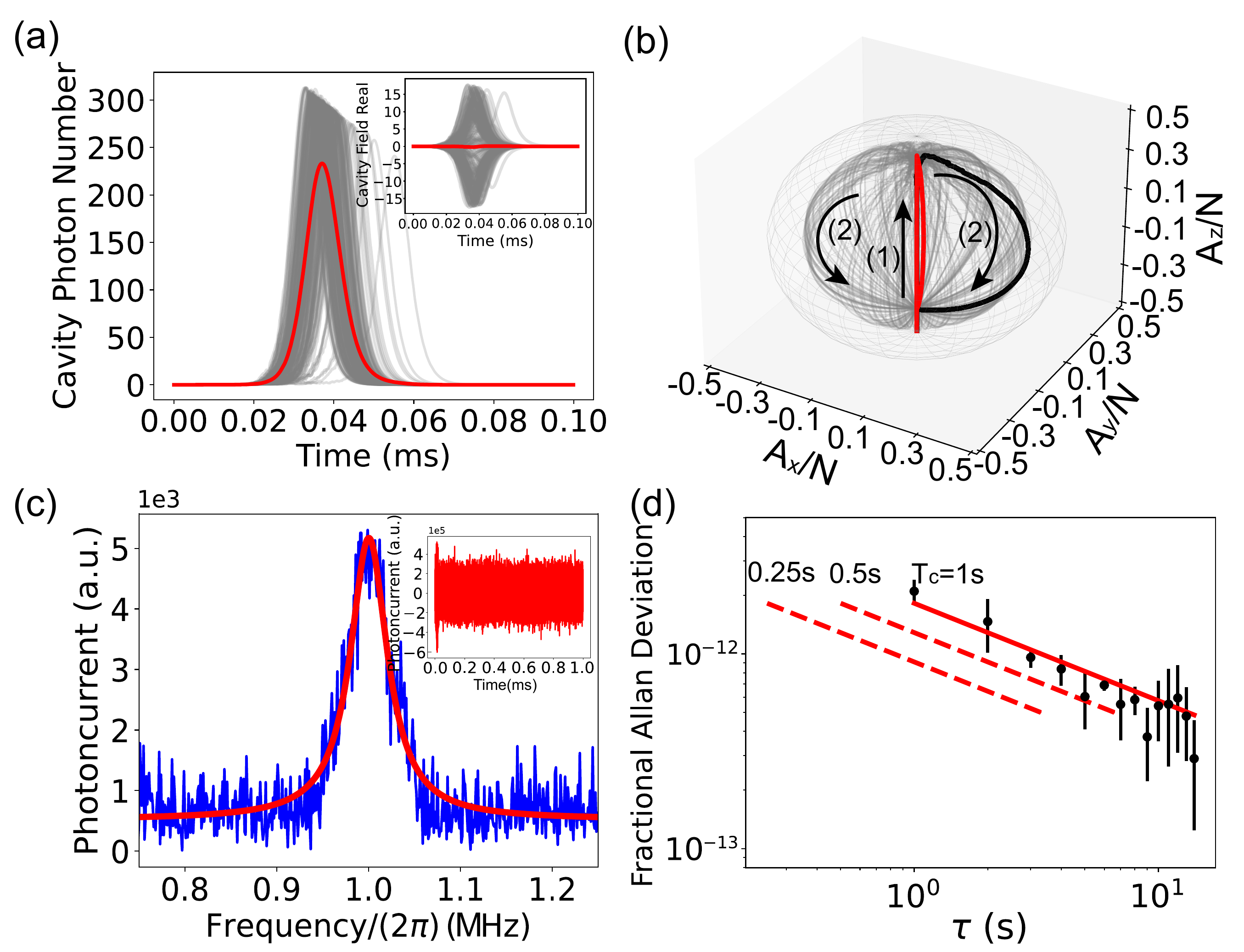}
    \caption{\label{fig:hd}  Active frequency measurement with heterodyne detection of superradiant pulses from incoherently pumped calcium atoms. Panel (a,b) show the trajectories of the intra-cavity photon number (a) and the real part of the field amplitude (inset of a), and the collective spin vectors (b) in different simulations (gray lines) and their averages (red solid lines). In the panel (b), the labels (1),(2) mark the direction of increasing time. Panel (c) shows the power spectrum obtained by the Fourier transform of the heterodyne signal  (inset), and the fitting of the spectral peak with a Lorentzian function (red solid line). Panel (d) shows the fractional frequency difference as function of the measurement time. Here, we consider the systems with $N=5\times10^4$ atoms, and the incoherent pumping pulse with a rate $20$ kHz and duration $20$ $\mu$s. Other parameters are given in Tab.~\ref{tab:params}. }
    \label{fig:measurement}
\end{figure}

In  Fig. \ref{fig:hd} we show our simulations of the system dynamics in the presence of heterodyne detection. Fig. \ref{fig:hd}(a) shows that the intra-cavity photon-number shows large variations in different simulations, but the average result reproduces the deterministic dynamics [Fig. \ref{fig:pulses}(a)]. Importantly, the intra-cavity  field amplitude develops non-vanishing values in individual simulations, while the average over many  simulations vanishes, see the inset of Fig. \ref{fig:hd}(a) and  Fig. \ref{fig:Cavity Field}(b). Since the field amplitude couples with the atomic coherence, the collective spin vector components $A_x,A_y$ also acquire finite values, and the collective spin vector departs from the z-axis and rotates on a spherical surface inside the outermost sphere, see Fig. \ref{fig:hd}(b), while the average over many trajectories recovers the deterministic dynamics along the z-axis [Fig. \ref{fig:pulses}(d)].  The quantum measurement backaction is the causes of a spontaneous breaking of the U(1) phase symmetry, which is often imposed as an ansatz in the semi-classical theory of superradiance from fully excited atoms~\citep{JCMacGillivray,DPolder}.

Fig.~\ref{fig:hd}(c) shows the calculated power spectrum of the heterodyne detection obtained by the Fourier transform of the simulated photoncurrent (inset). We see that the envelope of the rather noisy current decays dramatically in short time and then saturates, which actually follows the real part of the field amplitude [inset of Fig.~\ref{fig:hd}(a)]. The noisy power spectrum shows a peak around $2\pi$ MHz, i.e. the frequency detuning of the reference laser and the optical cavity, with a width around $2\pi\times 28$ kHz, and can be well fitted by a Lorentzian function. In Fig. S7 of the SI, we have examined the dependence of the fitted frequency, width and  signal-to-noise ratio on the time span of the Fourier transform. We find that the former two quantities get more precise due to the improved frequency resolution, while the latter decreases because of the white noise contribution for time longer than the superradiant signal. Furthermore, by following the protocol used in Refs. ~\citep{MANorcia2018,YZhang2021-1}, we combine individual short time simulations to form simulations for longer measurement time, and we calculate the fractional Allan deviation $\sigma(\tau)$ as function of measurement time $\tau$, see Fig. \ref{fig:hd}(d). If the single measurement lasts for one second ($T_c=1$ s), including possibly the time to reload the atoms,  the Allan deviation scales as $\sigma(\tau)\approx 1.8\times 10^{-12}/\sqrt{\tau/s}$ (rightmost line). If the time for a single measurement is reduced   ($T_c=0.5,0.25$ s), the deviation becomes also correspondingly reduced (lower dashed lines).

Finally, we would like to recall the connection between the power spectrum obtained by recording or simulating the heterodyne detection signal and the one calculated deterministically by the two-time field correlation function and the quantum regression theorem~\citep{PMeystre}.
These are, indeed, equivalent methods to obtain the spectrum. The original arguments by Glauber for the role of the time and normally ordered field correlation functions~\citep{JRGlauber} are precisely rooted in the annihilation of a photon from the emitter system, accompanying the photodetection event~\citep{LMandel}. In homodyne and heterodyne detection, this annihilation process mainly affects the local oscillator field and the back action on the emitter system amounts to an infinitesimal change of the state, proportional to the stochastic component of the measured signal~\citep{HWWiseman}. This, in turn, causes a finite correlation between signal values at different times, as observed in our simulations and in the inferred Fourier spectra. While the quantum regression theorem yields the average correlation functions and hence spectra, it cannot be simply applied to reproduce the outcome of fitting individual spectra by a Lorentzian to sample and analyze candidate values of the frequency.

\section{Discussions and Conclusions}

In summary, we have investigated the superradiant pulses of incoherently pumped calcium atoms trapped in an optical lattice, and demonstrated that these pulses are caused by incoherent de-excitation dynamics among Dicke states. Thus, the mechanism seems to be fundamentally different from the coherent rotation of a collective spin vector as explored in Ref.~\citep{MANorcia2018}. Still, stochastic simulations show that due to quantum measurement backaction, heterodyne detection of the superradiant pulses establishes coherence in the intra-cavity field and the atoms, and the phase stability of this coherence leads to the consistent peaks in the calculated power spectra. 


In this article, we focused on calcium atoms incoherently pumped to the short-lived excited state, as examined in the experiment~\citep{TLaske}. As a result, the calculated power spectrum is rather broad and the resulting performance of the frequency measurement is  comparable with the one of commercial Cesium clocks~\citep{ABauch}. However, by exploring the long-lived excited state of the calcium atoms~\citep{AVTaichenachev} or other atoms like strontium atoms~\citep{MANorcia2018}, we expect that the performance of the frequency measurement with incoherently pumped atoms can be improved by orders of magnitude. Furthermore, by applying continuous atomic pumping, one should be able to carry out the frequency measurement with steady-state superradiance~\citep{DMeiser2009,JDBohnet2012}.


\begin{acknowledgments}
This work was supported by the National Natural Science Foundation of China through the project No. 12004344,  62027816, and Henan Center for Outstanding Overseas Scientists Project No. GZS201903, as well as the Danish National Research Foundation through the Center of Excellence for Complex Quantum Systems (Grant agreement No. DNRF156).
\end{acknowledgments}

\appendix
\renewcommand\thefigure{A\arabic{figure}}
\renewcommand\thetable{A\arabic{table}}
\renewcommand{\thepage}{A\arabic{page}}
\renewcommand{\bibnumfmt}[1]{[S#1]}
\setcounter{figure}{0}

\section{Calcium Atomic Levels and System Parameters}

\begin{figure}[!htp]
\begin{centering}
\includegraphics[scale=0.23]{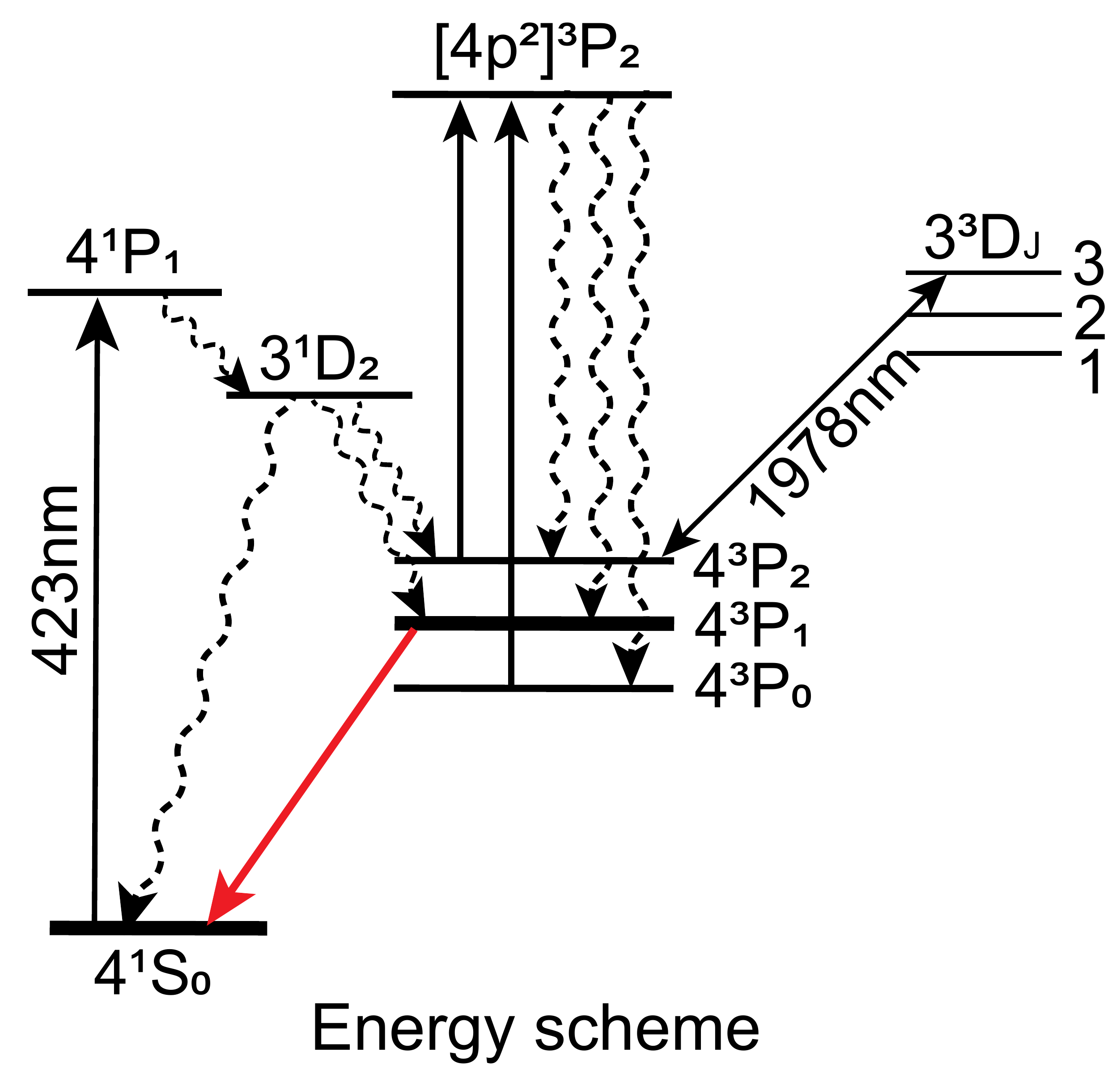}
\par\end{centering}
\caption{\label{fig:scheme} Schematic of calcium atoms levels. The solid and dashed arrows indicate the optical pumping and the spontaneous decay, while the red solid line labels the collective decay, leading to the superradiant pulses.}
\end{figure}

In this Appendix, we describe the schematic of calcium atomic levels, see Fig. \ref{fig:scheme}, and the parameters used in our simulations, see Tab. \ref{tab:params}.

In the experiment~\citep{TLaske}, the calcium atoms were initially transferred from the ground state $4^1{\rm S}_0$ to the $4^3{\rm P}_2$ excited state by laser excitation ($423$ nm) to the $4^1{\rm P}_1$ state and subsequent decay via the $3^1{\rm D}_2$ state. Then, the atoms were trapped in a magneto-optical trap using the $4^3{\rm P}_2-3^3{\rm D}_3$ transition, and were further transferred to the $4^3{\rm P}_1$ state in an optical lattice by pumping to the $[4{\rm p}^2]^3{\rm P}_2$ excited state  and the subsequent decay.  Finally, these excited atoms decay collectively to the ground state by the weak coupling with the optical cavity.

\begin{table}[!htp]
\begin{centering}
\begin{tabular}{|c|c|c|c|}
\hline 
$\omega_{c}$ & ${2\pi\times456.6}$ THz& $\eta$ & $2\pi\times20$ kHz\tabularnewline
\hline 
$\kappa$ & $2\pi\times2.26$ MHz & $\chi$ & $0.1$ Hz\tabularnewline
\hline 
$\omega_{a}-\omega_{c}$ & $0$ & $N$ & $5\times10^{4}$\tabularnewline
\hline 
$g$ & $6.53$kHz & $\omega_{l}-\omega_{c}$ & $2\pi$ MHz\tabularnewline
\hline 
$\gamma$ & $2\pi\times0.38$ kHz & $\xi$ & $0.12$\tabularnewline
\hline 
\end{tabular}
\par\end{centering}
\caption{System parameters for the simulations in the main text and the Appendix. \label{tab:params}}
\end{table}

The system parameters are shown in Tab.~\ref{tab:params}. The optical cavity has a frequency $\omega_c= 2\pi\times456.6$ THz, and a photon damping rate $\kappa = 2\pi\times2.26$ MHz. The calcium atoms couple resonantly with the cavity mode, i.e. $\omega_a=\omega_c$ , with a strength $g=6.53$ kHz. The atoms have  a decay rate $\gamma=2\pi\times 0.38$ kHz, a dephasing rate $\chi =0.1$ Hz, and an incoherent pumping rate $\eta = 2\pi \times 20 $ kHz. In our simulations, the reference laser is detuned from the cavity mode by $\omega_l-\omega_c = 2\pi$ MHz, and the detector efficiency is $\xi =0.12 $.

\section{Julia Codes to Solve Quantum Master Equation}

\begin{figure}
\begin{centering}
\includegraphics[scale=0.75]{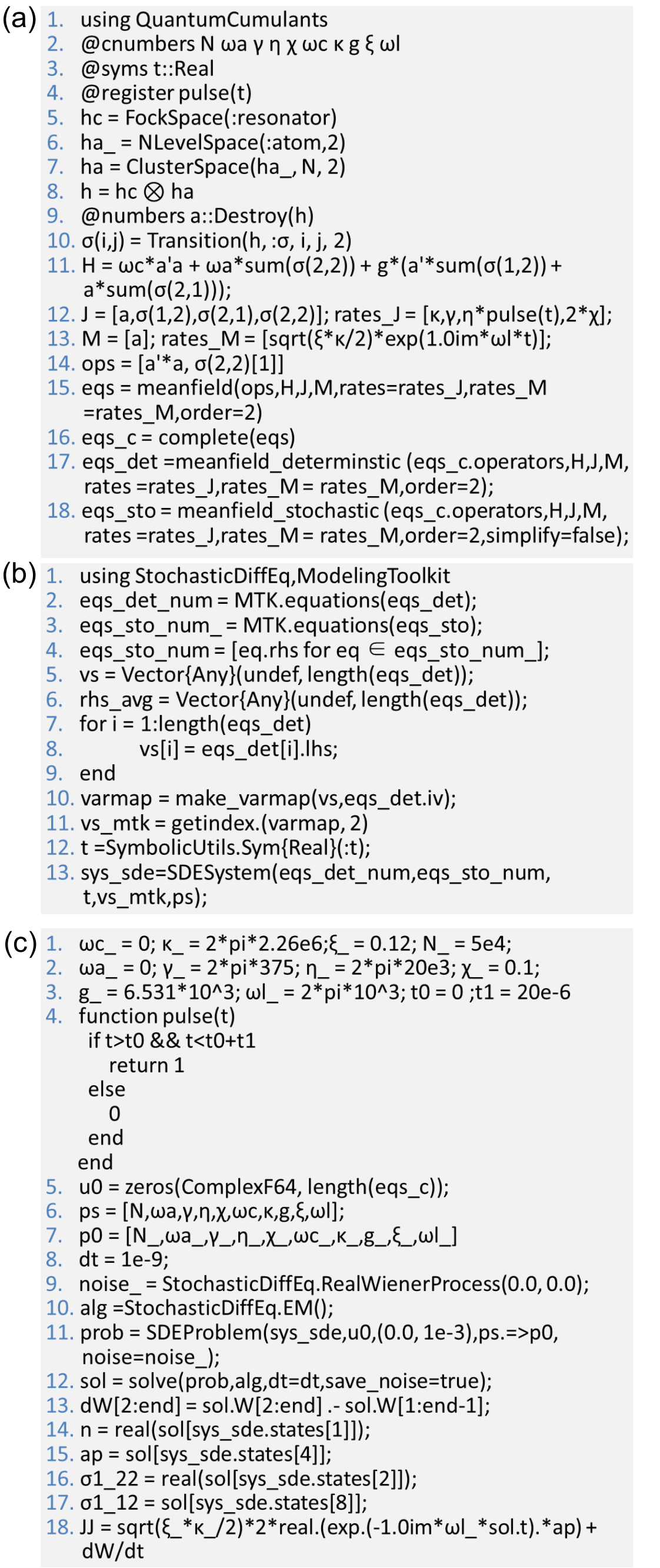}
\par\end{centering}
\caption{\label{fig:code}Julia codes to  solve the stochastic master equation in the main text within the mean-field approach. Panel (a) shows the codes to derive the symbolic equations for the mean-field quantities to second order. Panel (b) shows the codes to convert the symbolic equations to numerical ones. Panel  (c) shows the codes to solve the numerical stochastic equations.  } 
\end{figure}

In this Appendix, we describe the Julia codes to solve the quantum master equation,  see Fig. \ref{fig:code}. Fig. \ref{fig:code}
(a) shows the codes to derive the mean-field equations in second order. In lines 1-8, we import the QuantumCumulants.jl package, define the symbols for the complex variables to be used later, define the time and the pump pulse, define the Hilbert space for the two-level atoms and $N$ identical two-level atoms, define the quantized harmonic oscillator Fock space for the optical cavity mode, define the product Hilbert space for the atoms-cavity mode system. In lines 9-16, we define the photon annihilation operator $a$ (and its conjugation $a'$) and the transition and projection operators for the atoms, define the system Hamiltonian, derive the mean-field equations for three specific operators, define lists of operators
and rates to construct the Lindblad superoperators, define lists of operators and rates to construct the measurement backaction contribution, derive the mean-field equations for three specific operators, and finally complete the equations with remaining mean-field quantities, to obtain a complete set of mean-field quantities. 

In lines 17 and 18, we derive the contributions of the deterministic master equations and the measurement backaction. Here, the function ``meanfield\_deterministic'' is same as the ``meanfield'' function in the QuantumCumulants.jl package, and we have defined a new function ``meanfield\_stochastic'', where we evaluate the contributions of the measurement backaction to the mean-field equations. Note that when deriving the mean-field equations, we work in a frame rotating with the cavity frequency $\omega_{c}$ by setting it to zero. 

Fig. \ref{fig:code} (b) shows the Julia codes to define the system of stochastic differential equations. In the 1st line, we import the ``StochasticDiffEq.jl'' and ``ModelingToolkit.jl'' package, which provides the solvers to the stochastic differential equations,
and a modeling framework for the high-performance symbolic-numerical computation. In lines 2-4, we define a model for the deterministic and stochastic part of the mean-field equations. In lines 4-11, we extract the left-hand side of the mean-field equations, and define them as functions of time. In lines 12-13, we define a symbol for time argument and the system of the stochastic differential equations. 

Fig. \ref{fig:code} (c) shows the Julia codes to solve numerically the mean-field equations. In lines 1-5, we specify values for the involved parameters and the pump pulse. In line 6, we define the initial value for the mean-field quantities. In lines 7 and 8, we define the lists of parameters and their values. In line 9, we define time steps.In lines 10 and 11, we define the noise and integration method. In lines 12 and 13,we define the stochastic differential equations and solve them numerically with the Euler-Maruyama method.In lines 14-19, we calculate the shot-noise,the cavity photo number, the field amplitude,the population, the atom-atom correlations. Here we can also get the cavity field amplitude, atomic coherence, atomic-photon correlation, atom-atom correlation, and then compute the heterodyne detector current, based on which we can calculate the quantum numbers of the Dicke state and the collective spin vector.

\begin{figure}
\begin{centering}
\includegraphics[scale=0.56]{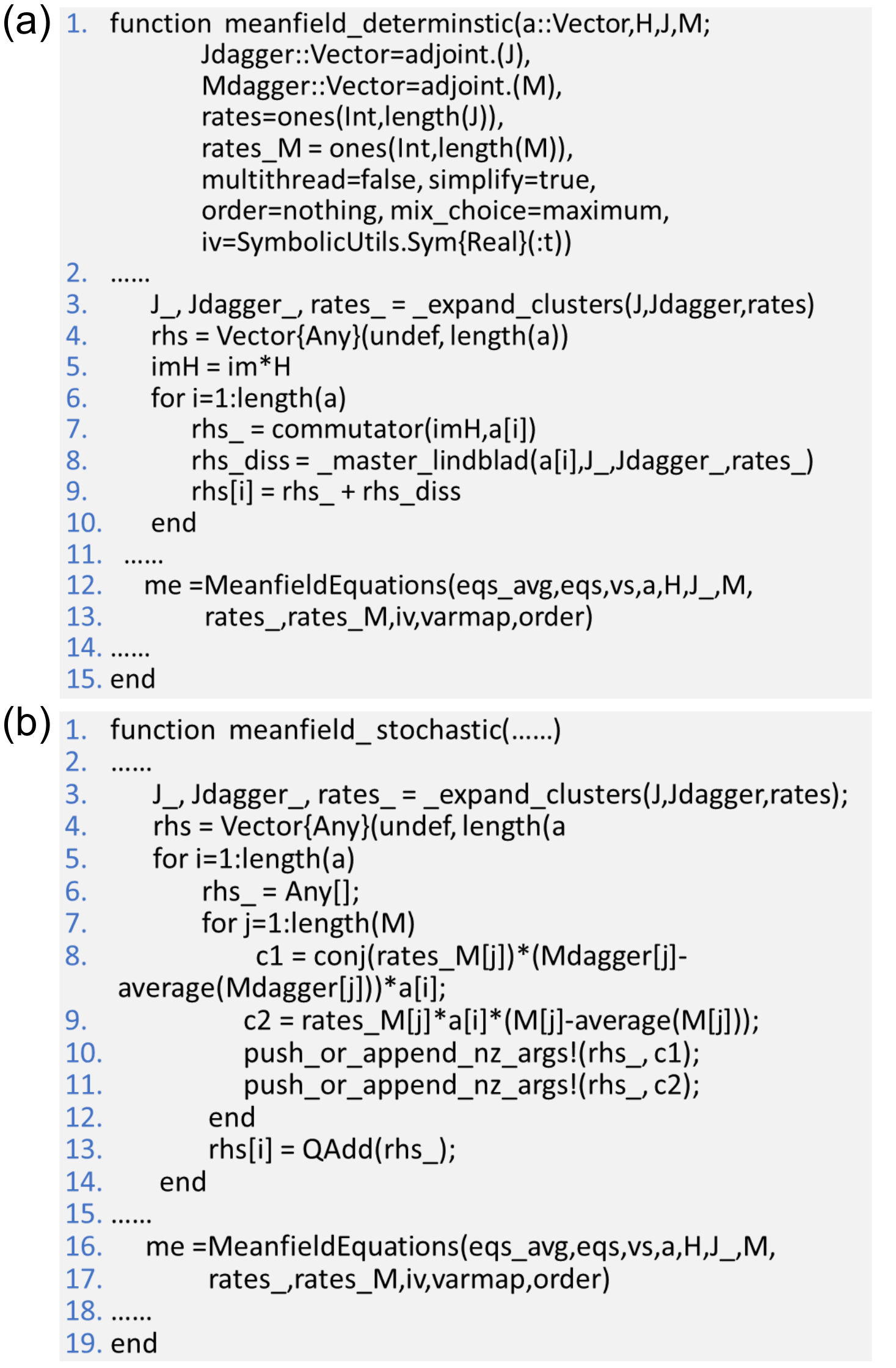}
\par\end{centering}
\caption{\label{fig:mf_det_sto} Partial codes of the " meanfield\_deterministic" function (a) and the " meanfield\_stochastic" function (b). } 
\end{figure}

While the QuantumCumulants.jl package is designed to solve the standard, deterministic quantum master equation, we have modified it to solve the stochastic master equation. In the modification, the most significant change is to define a new function "meanfield\_stochastic" to derive the dependence of the mean-field quantities due to the measurement backaction. 

In Fig. \ref{fig:mf_det_sto}(a), we show partial codes for the "meanfield\_deterministic" function which resembles the "meanfield" function in the QuantumCumulants.jl package. The first line defines the function with the necessary parameters. The first two parameters are the list of operators "a" to define the meanfield quantities, the system Hamiltonian "H". Lines 2 and 3  are the list of operators "J,M" to define the Lindblad and stochastic term. Lines 4 and 5 define the conjugation "Jdagger,Mdagger" of "J,M", and the next two parameters are the rates "rates,rates\_M" to complement "J,M". Lines 6-8 define remaining parameters. Among them, "order,iv" are the order of the mean-field approach, and the symbol for the time parameter. Line 10 translates the parameters "J,Jdagger" in the full Hilbert space to the reduced space, which exploits the permutation symmetry due to the assumption of identical particles. Line 11 defines the array of operators with the same length as the list "a". Line 12 defines for convenience "imH" as the multiplication of the imaginary sign with the system Hamiltonian. In lines 13-17, we compute the time dependence due to the communication relation of "a" and "imH", and due to the Lindblad term, and we form the equations for the operators in the list "a". Later, we transfer these equations into "eqs\_avg" for the mean-field quantities by using the "average()" function (not shown). The 19th line defines the object "me" to represent the derived mean-field equations. One should note that in this function the lists "M,Mdagger,M\_rates" for the stochastic terms are used only for the definition of the object "me". 

In Fig. \ref{fig:mf_det_sto}(a), we show partial codes for the "meanfield\_stochastic" function. Only lines 5-14 differ from the ones in the "meanfield\_deterministic" function. In this code, we define an empty list "rhs\_", and then enumerate every term in the list "M", and compute the dependence "c1,c2" due to the corresponding stochastic terms, and push them into the list "rhs\_", and finally add the terms in the list to form the equations for the operator in the list "a". Here, the most important step is to introduce the mean-field terms through the "average()" function into the equations for the operators.  In line 16, we define again the object "me" to represent the derived mean-field equations. One should note that in this function the lists "J,Jdagger,J\_rates" for the deterministic terms are used only for the definition of the object "me".  

The modified code is suitable for one particular stochastic master equation, where the single cavity mode is treated as a component of the system. However, in some cases, one may eliminate the cavity mode, and deal with the stochastic master equation only for the atoms, where the cavity mode or radiation field enters into the equation through parameters which gauge the strength of the collective decay and measurement backaction. In principle, the same technique as presented above can be applied to solve this equation. However, the actual implication might need further consideration, and is beyond the scope of the current work.

\section{Supplemental Results}
In this Appendix, we provide extra results to complement the discussions in the main text.

\begin{figure}
\begin{centering}
\includegraphics[scale=0.24]{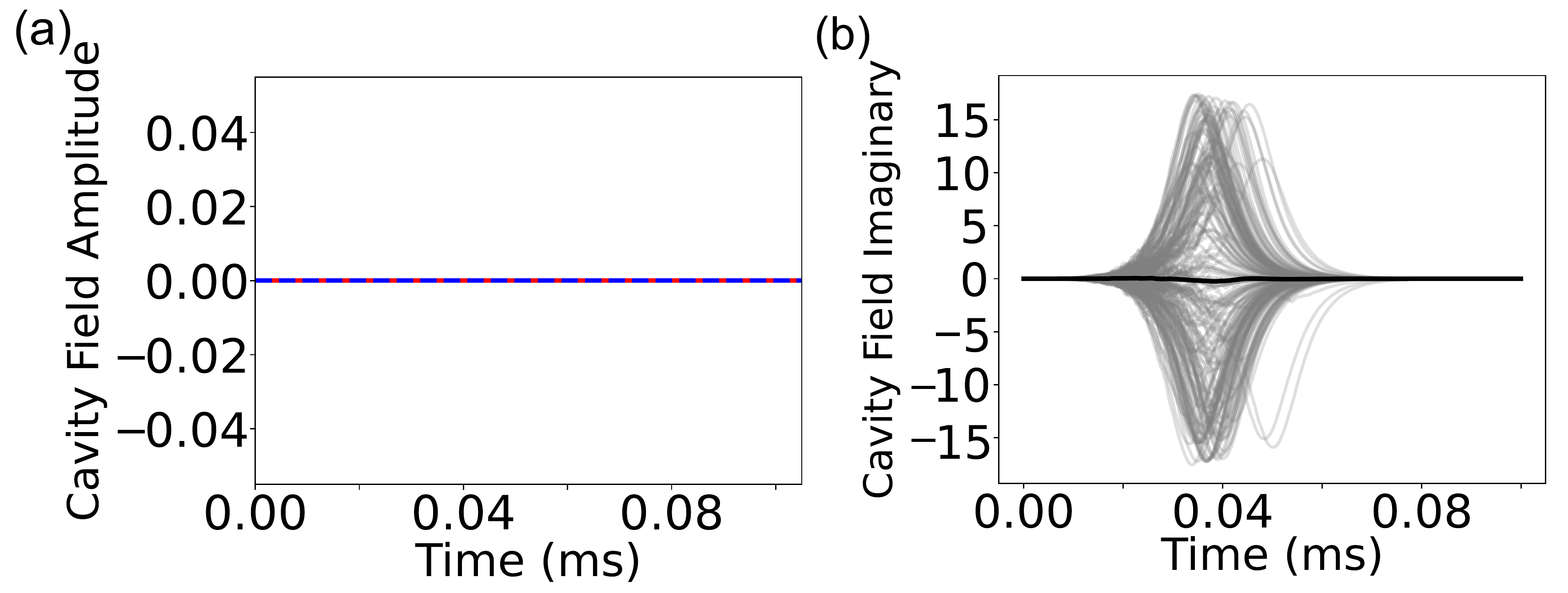}
\par\end{centering}
\caption{\label{fig:Cavity Field} Supplementary results for the intra-cavity field amplitude. Panel (a) displays the real (red  solid line) and imaginary (blue dashed  line) parts of the intra-cavity field amplitude (both vanish) in the absence of the heterodyne detection. Panel (b) shows the imaginary part (gray lines) and its mean (black line) of the intra-cavity field amplitude for $200$ simulations in the presence of heterodyne detection. The results for the real part are shown in the inset of Fig. 3(a) of the main text. }
\end{figure}

\subsection{Intra-cavity Field Amplitude}

 In the main text, we indicate that the intra-cavity field amplitude vanishes in the absence of the heterodyne detection, while the quantum measurement backaction during the heterodyne detection introduces a non-vanishing mean field amplitude.  These points are clearly illustrated in Fig. \ref{fig:Cavity Field}. We also note that the intra-cavity field amplitude varies dramatically in individual simulations, but their average is very close to zero, recovering the dynamics in the absence of the detection.

\begin{figure}
\begin{centering}
\includegraphics[scale=0.25]{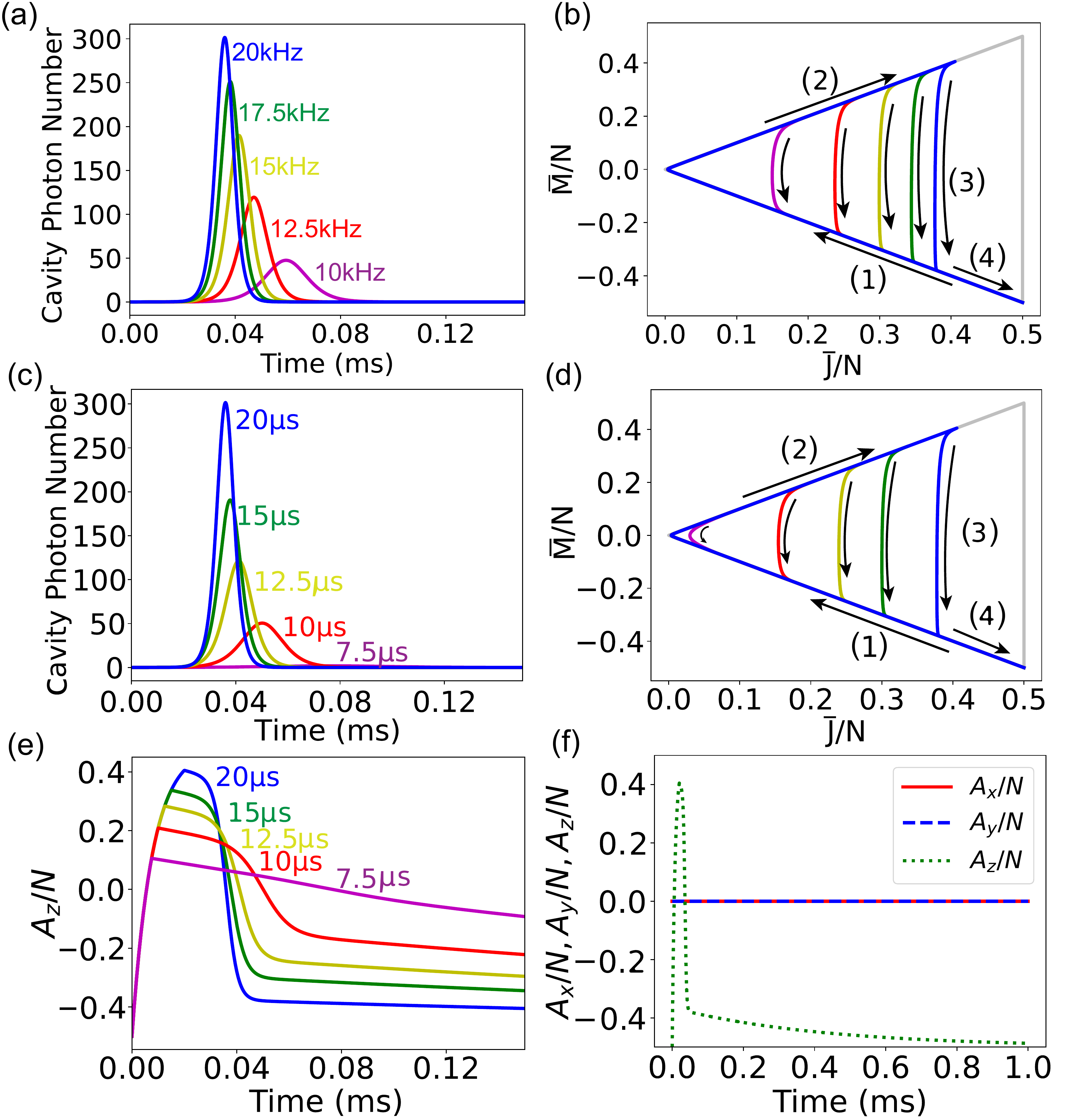}
\par\end{centering}
\caption{\label{fig:pump time} Supplementary results for the superradiant pulses with incoherently pumped atoms. Panels (a,b) display the intra-cavity photon number (a), the averaged Dicke state quantum numbers  $\overline{J},\overline{M}$ (b) for systems with different pumping rates and fixed pumping time $t=20$ ${\rm \mu s}$. Panel (c,d) show similar results for a fixed pumping rate $\eta =2\pi \times 20$ kHz and different pumping times.  Panel (e) displays the collective spin vector component $A_z$ (normalized by the number of atoms $N$)  for the systems with different pumping time, while the situation of different pump rates is similar (not shown). Panel (f) shows the variation of  three collective spin vector components in a longer time range. In these simulations, we consider the system with $N=5\times 10^4$ atoms, and the incoherent pumping with a rate $ \eta =2\pi\times 20$ kHz and a duration $t=20$ $\mu$s (for the panel f).}
\end{figure}

\subsection{Extra Results for System with Incoherently Pumped Atoms}

In the  main text, we discussed superradiant pulses from the incoherently pumped atoms, and we showed the system dynamics for different numbers of atoms in Fig. 2. We supplement these results with superradiant pulses for different pumping rates and durations, see Fig. \ref{fig:pump time}(a-e). We find that as the pumping duration or rate increases, the superradiant pulses appear earlier and they are stronger and more peaked [Fig. \ref{fig:pump time}(a,c)]. These results are caused by pumping the atoms to the Dicke states with higher excitation (larger $\overline{J}$ and $\overline{M}$) and the subsequent vertical decay  through the Dicke states with large $\overline{J}$ [Fig. \ref{fig:pump time}(b,d)], indicating also a larger coupling with the optical cavity mode. As the pumping time increases, the collective spin vector component $A_z$ (normalized by the number of atoms $N$) attains higher value [Fig. \ref{fig:pump time}(e)], and the decay is more steep, which is consistent with the change in the Dicke state space. Note that the other vector components $A_x,A_y$  always vanish [Fig. \ref{fig:pump time}(f)].

In the above discussion and in the main text, we focused on the superradiant dynamics in the short time range. After this collective decay dynamics, the atomic ensemble ends up in states on the lower boundary of the Dicke state space and eventually, under the influence of individual decay, the atomic ensemble converges along the lower boundary to the general ground state in the lower right corner [Fig.~\ref{fig:pulses}(b)]. Accompanying this dynamics, the collective spin vector component  $A_z/N$  decreases slowly to the value $-1/2$, reaching eventually the south pole of the collective spin sphere.

\begin{figure}
\begin{centering}
\includegraphics[scale=0.25]{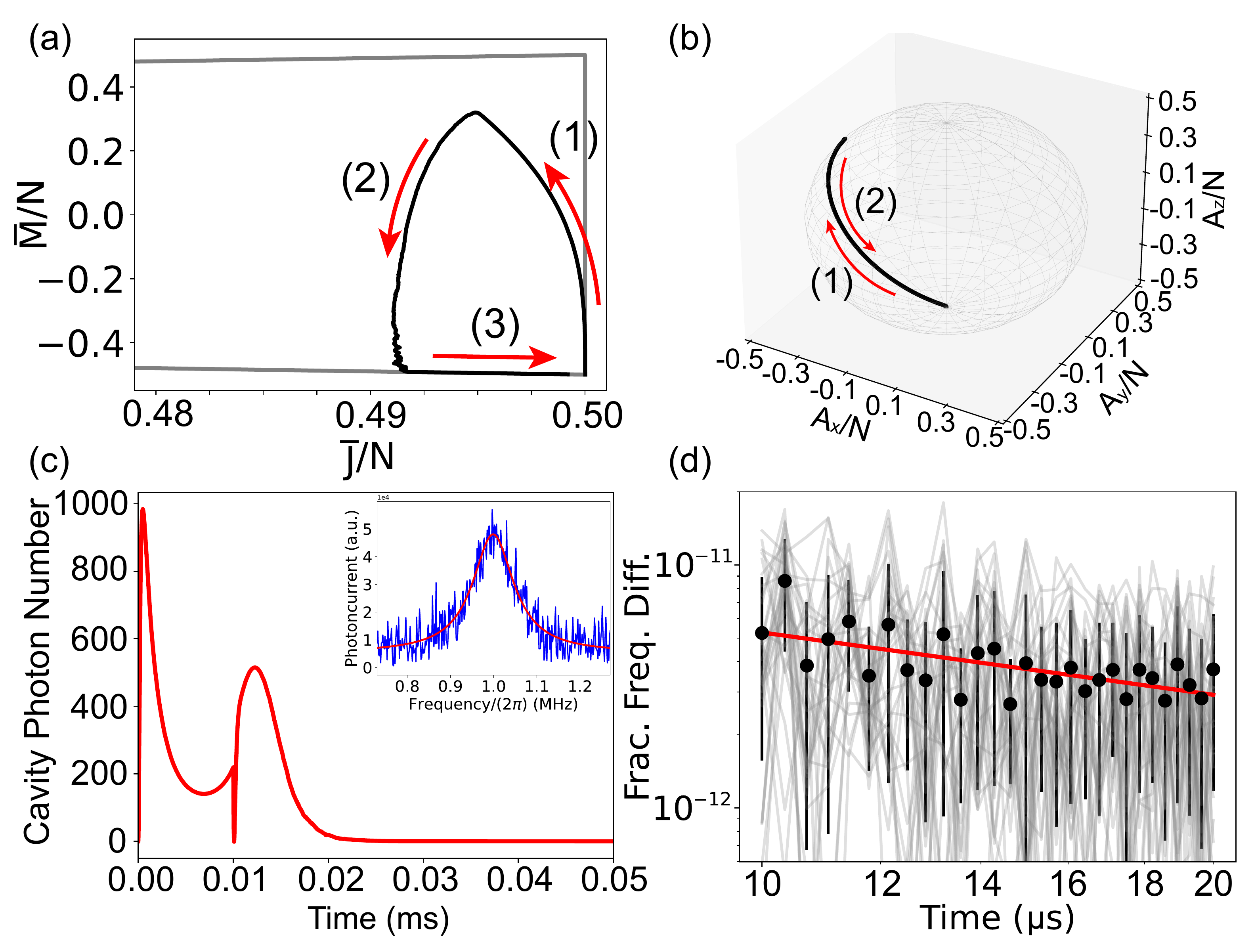}
\par\end{centering}
\caption{\label{fig:Coherent pump} { Heterodyne detection-based frequency measurement with the superradiant pulses from coherently pumped calcium atoms. Panels (a,b) show the dynamics of  the Dicke state numbers and the collective spin vector during the laser driving and superradiant decay period. Panel (c) shows the corresponding dynamics of the intra-cavity photon number, and the simulated power spectrum (inset). Panel (d) shows the calculated fractional frequency different as function of the measurement time. In the simulations, the driving laser is resonant with the cavity mode, i.e. $\omega_d=\omega_c$, and the driving strength is  $\Omega =2.9\pi\times10^4$ $\sqrt{\rm Hz}$ for $0.01$ ms. } }
\end{figure}

\subsection{Heterodyne Detection of the System with Coherently Pumped Atoms \label{sec:MeaCoh}}

In the main text, we found that the quantum measurement backaction plays an essential role in establishing the heterodyne detection record for the superradiant pulses from the incoherently pumped calcium atoms. This situation seems to be different from systems with coherently pumped atoms, as investigated with strontium-87 atoms in the experiment~\citep{MANorcia2018} and theory~\citep{YZhang2021-1}.  

To illustrate the above point further, we investigate the case of coherently pumped calcium atoms. To simulate this system, we first complement the quantum master equation $\partial_t \hat{\rho}$  with the laser driving Hamiltonian $\hat{H}_d =  \sqrt{\kappa/2} \Omega (e^{i\omega_d t}\hat{a} + h.c.)$, where  $ \sqrt{\kappa/2}$ is the assumed amplitude transmission coefficient of the left mirror, $\omega_d$  is the frequency of the driving laser, $\Omega$ controls the strength of the laser driving. Following the similar codes as shown in Fig. \ref{fig:code}, we solve the modified master equation with the QuantumCumulants.jl package, and obtain the results shown in Fig. \ref{fig:Coherent pump}. 

During the coherent driving, the atomic ensemble moves vertically from the lower-right corner to the point near the upper-right corner of the Dicke state space [Fig. \ref{fig:Coherent pump}(a)],  and the collective spin vector rotates around the x-axis to some point near to the north pole [Fig. \ref{fig:Coherent pump}(b)]. During the superradiant emission, the atomic ensemble decays vertically back to the lower-right corner of the Dicke state space [Fig. \ref{fig:Coherent pump}(a)], and the collective spin vector rotates reversely back to the south pole [Fig. \ref{fig:Coherent pump}(b)]. These results are dramatically  different from Fig. 2(b) and  3(b) in the main text.  The most significant differences are that the coherent dynamics involves states on the surface of the collective spin sphere, and Dicke states with larger values of $\overline{J}\approx N/2$.  Since the collective spin vector component $A_x$ is not zero during the superradiant phase, the cavity field amplitude $\left \langle \hat{a}^{\dagger} \right \rangle$ also attains non-vanishing values. Furthermore, we find that the dynamics of $\left \langle \hat{a}^{\dagger} \right \rangle$ and the intracavity photon number $\left \langle \hat{a}^{\dagger} \hat{a} \right \rangle$  are almost unperturbed by the back action of the heterodyne detection. In any case, the simulated power spectrum is similar to the case with the incoherently pumped atoms, and the obtained fractional frequency difference is of the same order [Fig. \ref{fig:Coherent pump}(c,d)]. Here, the peak in the power spectrum is slightly wider because the superradiant pulses are slightly shorter for the atoms not fully excited. All in all, these results indicate that  the quantum measurement backaction does not play a significant role in heterodyne detection of superradiant pulses from coherently pumped atoms.

\begin{figure}
\begin{centering}
\includegraphics[scale=0.27]{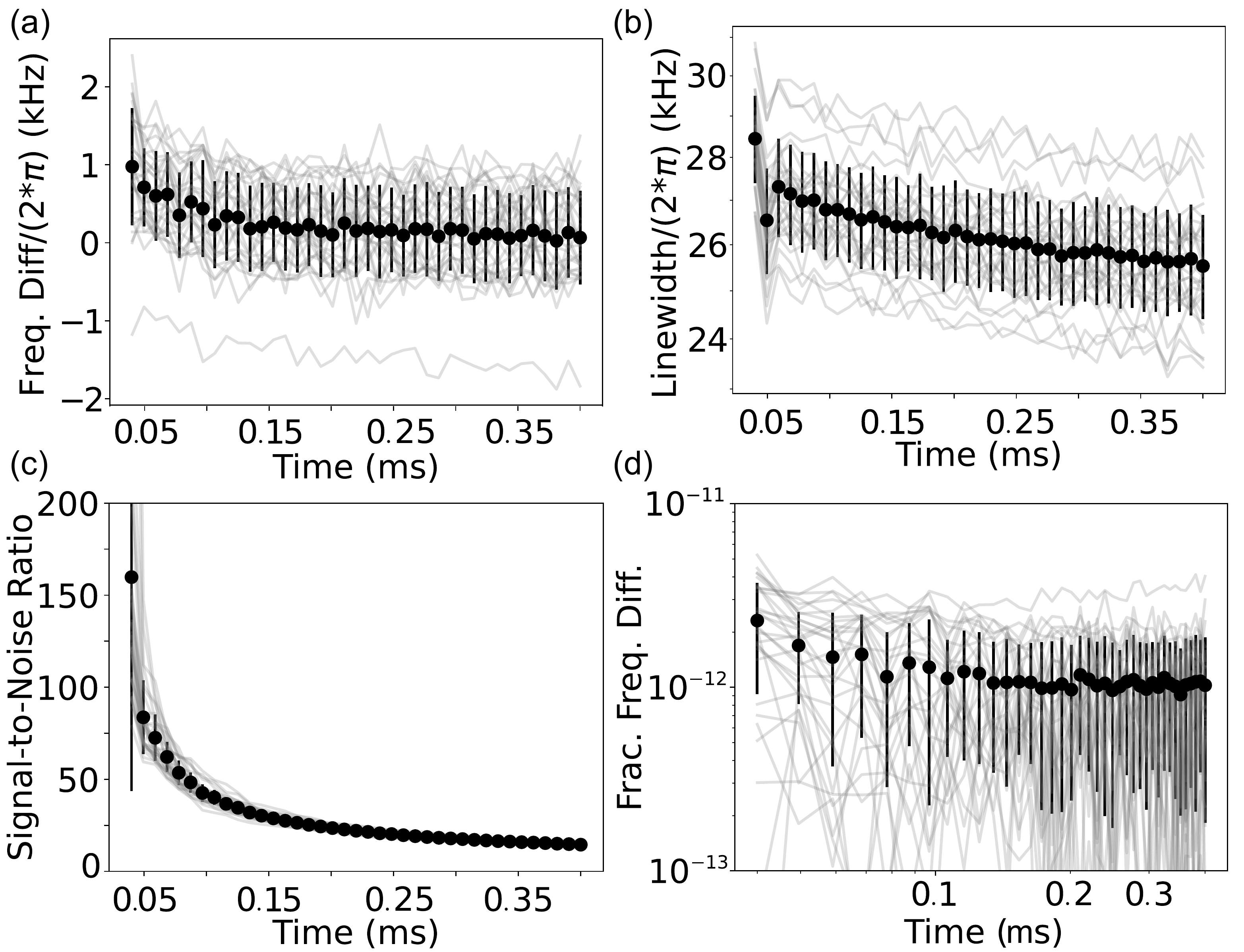}
\par\end{centering}
\caption{\label{fig:ffd}  Supplementary results for frequency measurements with superradiant pulses of the incoherent pumped calcium atoms.  Panel (a,b,c)  show the fitted frequency, linewidth, and   signal-to-noise ratio of the single peak power spectrum as function of the time span of the Fourier transform. Panel (d) shows the fractional frequency difference as function of the time span. The gray lines show the results of $200$ individual simulations, and the dots and bars show the mean and average of the results. }
\end{figure}

\subsection{Fractional Frequency Difference}

In Fig. 3(c) of the main text, we presented the dynamics of the photocurrent in the presence of heterodyne detection, and we obtained the power spectrum obtained by the Fourier transform of the current, which we fitted with a Lorentzian function.  In Fig. \ref{fig:ffd}, we complement this result by showing the fitted frequency $\omega_f$  (a), linewidth $\delta_f$  (b) and signal-to-noise ratio $r_f$ (c), as well as the calculated fractional frequency difference $d_f=(\omega_f -\omega_a)/\omega_a$ (d) as function of the time span $\tau$  of the Fourier transform.  The frequency $\omega_f$ and linewidth $\delta_f$ get more precise as the time span increases, because the frequency resolution of Fourier transform improves.
However, their error bar does not change much, since the time span considered here is longer than the signal (about $0.04$ ms). The signal-to-noise ratio $r_f$ decreases dramatically from about $150$ to about $10$ as the time span increases to $0.4$ ms. Since the superradiant signal lasts about $0.05$ ms, the heterodyne detection actually measures the white noise for most of the time, and this causes the reduction of   $r_f$.  Since the change of  $d_f$ is derived from $\omega_f$,  they behave similarly. The presented $d_f$ saturates at around  $1\times 10^{-12}$, which is comparable with that of the commercial microwave Cesium clocks \citep{ABauch}.

\end{document}